\documentclass[a4paper]{jpconf} \usepackage{graphicx} 
\begin{document}
\title{Modeling contact formation between atomic-sized gold tips via molecular dynamics}

\author{W Dednam$^{1,2}$, C Sabater$^{1,4}$, M A Fernandez$^{1}$, C Untiedt$^{1}$, J J Palacios$^{3}$, M J Caturla$^{1}$} \address{$ˆ1$
  Departamento de Fisica Aplicada, Universidad de Alicante,  San
  Vicente del Raspeig, E-03690 Alicante, Spain} \address{$ˆ2$
  Department of Physics, Science Campus, University of South Africa,
  Private Bag X6, Florida Park 1710, South Africa} \address{$ˆ3$ Departamento de
  Fisica de la Materia Condensada, Universidad Autonoma de Madrid,
  Cantoblanco, Madrid 28049, Spain} \address{$^4$ Huygens-Kamerlingh Onnes Laboratorium,
  Leiden University, Niels Bohrweg 2, 2333 CA Leiden, Netherlands} \ead{wdednam@gmail.com}

\pdfoutput=1

\begin{abstract}
The formation and rupture of atomic-sized contacts is modelled by means of molecular dynamics simulations. Such nano-contacts are realized in scanning tunnelling microscope and mechanically controlled break junction experiments. These instruments routinely measure the conductance across the nano-sized electrodes as they are brought into contact and separated, permitting conductance traces to be recorded that are plots of conductance versus the distance between the electrodes. One interesting feature of the conductance traces is that for some metals and geometric configurations a jump in the value of the conductance is observed right before contact between the electrodes, a phenomenon known as jump-to-contact. This paper considers, from a computational point of view, the dynamics of contact between two gold nano-electrodes. Repeated indentation of the two surfaces on each other is performed in two crystallographic orientations of face-centred cubic gold, namely (001) and (111). Ultimately, the intention is to identify the structures at the atomic level at the moment of first contact between the surfaces, since the value of the conductance is related to the minimum cross-section in the contact region. Conductance values obtained in this way are determined using first principles electronic transport calculations, with atomic configurations taken from the molecular dynamics simulations serving as input structures.

\end{abstract}

\section{Introduction}

 The microelectronics industry faces a great challenge if it is to continue the trend that has characterized its success over the last few decades \cite{Moorelaw}. The size of the gate in a transistor is now approaching a few nm in dimension and new strategies must be devised to achieve smaller sizes. Nanotechnology is a promising field and potentially offers a solution to the above-mentioned problem. In this connection, one of the most important tools both in the development and in the characterization of systems at the nano-scale is the scanning tunnelling microscope (STM) \cite{STMref}. With the aid of an STM or mechanically controllable break-junction (MCBJ), another related technique, we can study the electronic transport in few-atom systems \cite{Agraitrev}. Such studies typically require two electrodes of the same metal and a piezo system to join and separate the electrodes with subatomic precision. By also connecting a battery across the electrodes, the current through the nano-contact can be measured, usually in units of the quantum of conductance $(G_0=2e^2/h)$ \cite{G0wess}.
 
\begin{figure}[ht]
\centering 
\begin{minipage}{18pc}
\includegraphics[width=\textwidth]{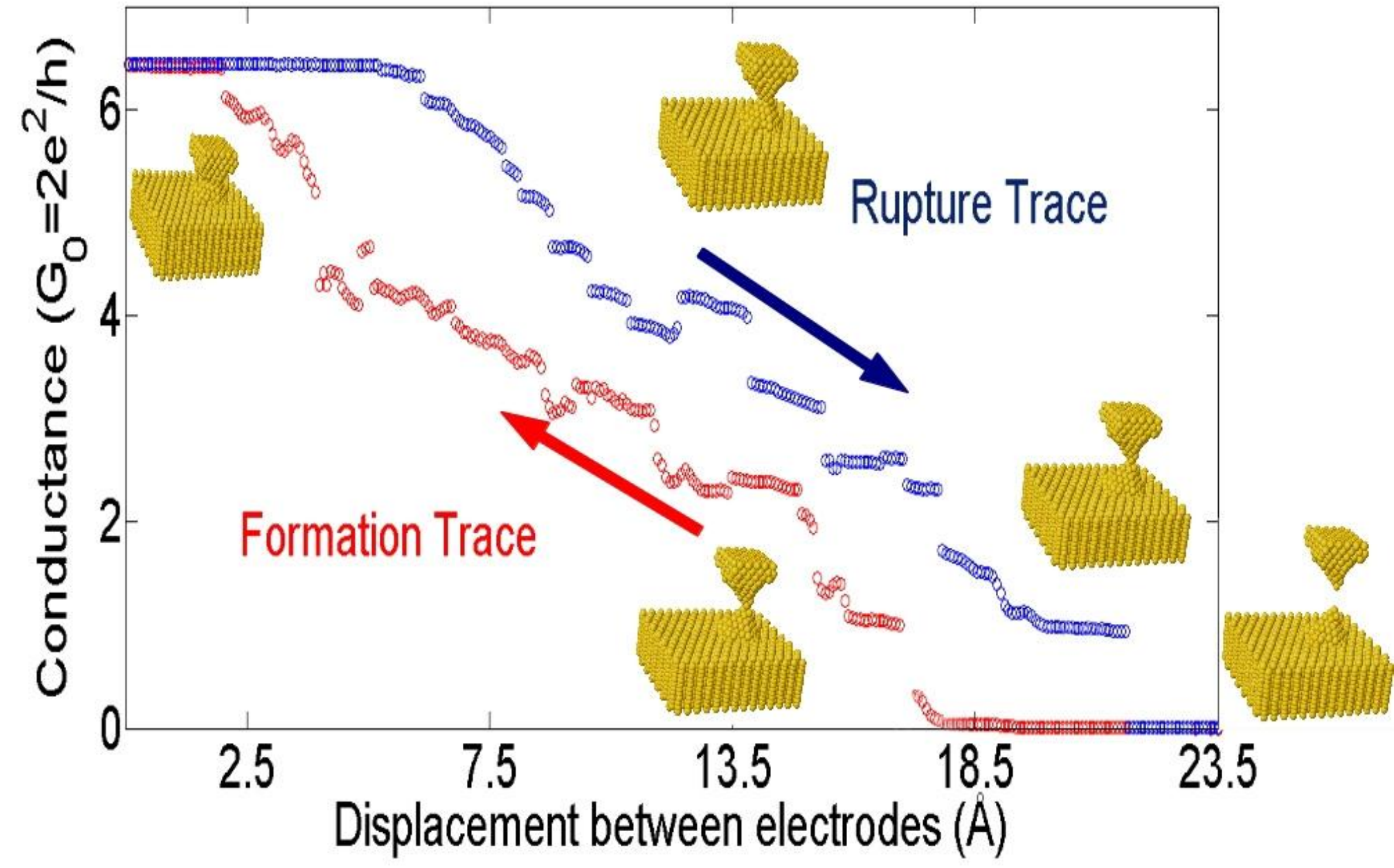}
\end{minipage}\hspace{1.5pc}
\begin{minipage}[b]{18pc}
\caption{(Colour online) Experimental conductance measurements recorded at 4.2 K using gold electrodes. The insets illustrate the geometry of the electrodes as contact is made or broken. \label{fig1}}
\end{minipage}
\end{figure}

The electronic transport through atomic-sized contacts can be interpreted from a plot of the measured conductance versus the displacement between the electrodes. Fig. \ref{fig1} shows experimental data for gold electrodes at $4.2$K; the blue curve corresponds to separation of the electrodes, and the red curve to the process of bringing them into contact. These curves are usually referred to as conductance traces. The insets in the figure help to visualise the contact formation and rupture processes. Moreover, the traces in the figure exhibit a jump in conductance \cite{UntiedtJC} that is close to 1 $G_0$ in value. In this case, the rupture trace corresponds to a transition from a monatomic contact to a complete loss of contact. In the case of the formation trace we observe a jump in the conductance that corresponds to a transition from tunnelling current to first atomic contact \cite{Gimtunatomic}.

A disadvantage of the above experimental techniques is the difficulty of imaging the electrodes at atomic resolution. Therefore, theoretical modelling can prove useful in obtaining more information about the geometry of the electrodes and the mechanism of contact formation. 

\section{Theory}

 Molecular dynamics (MD) simulations \cite{MD} based on empirical potentials can serve to model nano-contact formation and rupture. However, MD relies on interatomic potentials that are fitted to a limited number of parameters of the material. So, doubts remain as to the reliability of the potentials when systems are taken to situations far away from the range where the parameters have been fitted. Also, there is no description of the electrons in MD and it is thus not possible to obtain information such as the conductance of the nano-contacts.

On the other hand, DFT calculations \cite{DFT} have the advantage of being able to describe both the nuclei and the electrons in the system, with the use of certain approximations, but without the need of fitting parameters. Electronic transport calculations based on DFT permit the conductance to be obtained to a high level of accuracy in materials such as gold \cite{DFTcontacts}. However, these calculations are
computationally costly and systems of up to only a few tens of atoms can be conveniently handled.

We use a combination of these two methods to study the dynamical formation and breaking of gold nano-contacts and to calculate the conductance through the most probable structures at the moment of first contact. 

The MD code LAMMPS \cite{Lammps,Lammpsweb} is used to perform the dynamical calculations. The embedding atom potential \cite{EAM} for gold developed by Zhou \emph{et al.} \cite{EAMgold} serves to model the interactions between the atoms. This potential has been fitted to parameters such as lattice constants, elastic constants, bulk moduli, vacancy formation energies, sublimation energies, and heats of solution \cite{EAMgold}.

Note that it has been necessary to modify the LAMMPS source code to enable continuous breaking and making of nano-contacts. The simulation structure is divided along its length (z-axis) into evenly-spaced bins via a spatial averaging \emph{fix} that comes standard with LAMMPS. Following Ref. \cite{Mgwire}, the choice of bin size (height) is the interplanar distance between close-packed layers ($2.04$ \AA{} along  (001) and $2.3556$ \AA{} along (111)). Special functions have been added to the \emph{variable} subroutine in LAMMPS to find the bin containing the fewest atoms and the number of atoms in it. Each rupture-formation simulation is repeated over twenty cycles and paused every 10000 fs to determine the least-atom bin number and the number of atoms therein. 

\begin{figure}[ht]
\centering 
\begin{minipage}{18pc}
\includegraphics[width=\textwidth]{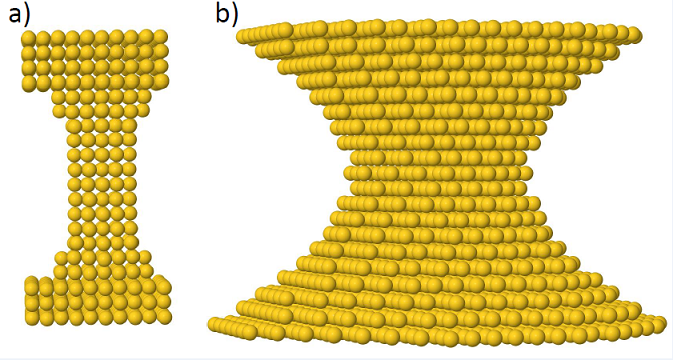}
\end{minipage}\hspace{1.5pc}
\begin{minipage}[b]{18pc}
\caption{(Colour online) Two of the initial configurations used in the simulations, left panel a) 525 atoms of gold, right panel b) 2804 atoms. Both are oriented along the (001) crystallographic direction in this case. \label{fig2}}
\end{minipage}
\end{figure}

System sizes ranging from 500 up to almost 3000 atoms and oriented along two different crystallographic directions, (001) and (111), have been considered. The initial configuration consists of a neck type structure (see Fig. \ref{fig2}). In the case of a narrow neck (e.g., Fig. \ref{fig2}(a)), the direction of motion during each rupture-formation cycle is reversed when there are a minimum of 5 atoms in the least-atom bin. In the case of wide neck (e.g., Fig \ref{fig2}(b)), on the other hand, the cycles are reversed when the least-atom bin contains 7 atoms. Each rupture or formation cycle is continued for a further 1000 fs after these conditions have been met to ensure that the contact has actually ruptured or that the least-atom bin indeed contains the predefined minimum number of atoms.

The first two top and bottom layers of all the simulated structures are frozen and displaced in opposite directions by $0.0041$ \AA{} and $0.0047$ \AA{} every 1000 fs for structures in the (001) and (111) crystallographic directions, respectively. The velocity of deformation in simulations (0.41 and 0.47 m/s along (001) and (111), respectively) are within the limits that can be computed by MD - about four orders of magnitude below the speed of sound in gold (3000 m/s) (See Ref. \cite{Speedsound} and the references therein). In order to keep the temperature constant throughout the calculation, a Nose-Hoover thermostat is applied \cite{Nose,Hoover}. 

For the formation cycles, a Fortran subroutine is used to calculate the minimum distance between the two opposing electrodes. Since the electrodes are allowed to separate to about 5 \AA{}, the atoms belonging to the top fragment are labelled "1" and those belonging to the bottom fragment "2". The first two distinctly labelled atoms ("1" and "2", respectively) to come within a distance halfway between first and second nearest neighbours during contact formation cycles, are then judged to have made contact first.

Subsequently, the conductance of selected structures at the moment of first contact - taken from the MD simulation trajectories - are computed by means of the electronic transport code ALACANT \cite{, Gaussian,Alacant}, using a minimal basis set that assigns only 1 orbital to each atom. It is well known that conductance values obtained from electronic transport calculations on few-atom nano-contacts, are determined by the few atoms in the minimum cross-section of the contact. Therefore, in order to save time on DFT calculations, we remove a few of the outer layers from each electrode.

\section{Results and Discussion}
\subsection{Molecular Dynamics Simulations}
\begin{figure}[htp!]
\centering
\includegraphics[width=0.49\textwidth]{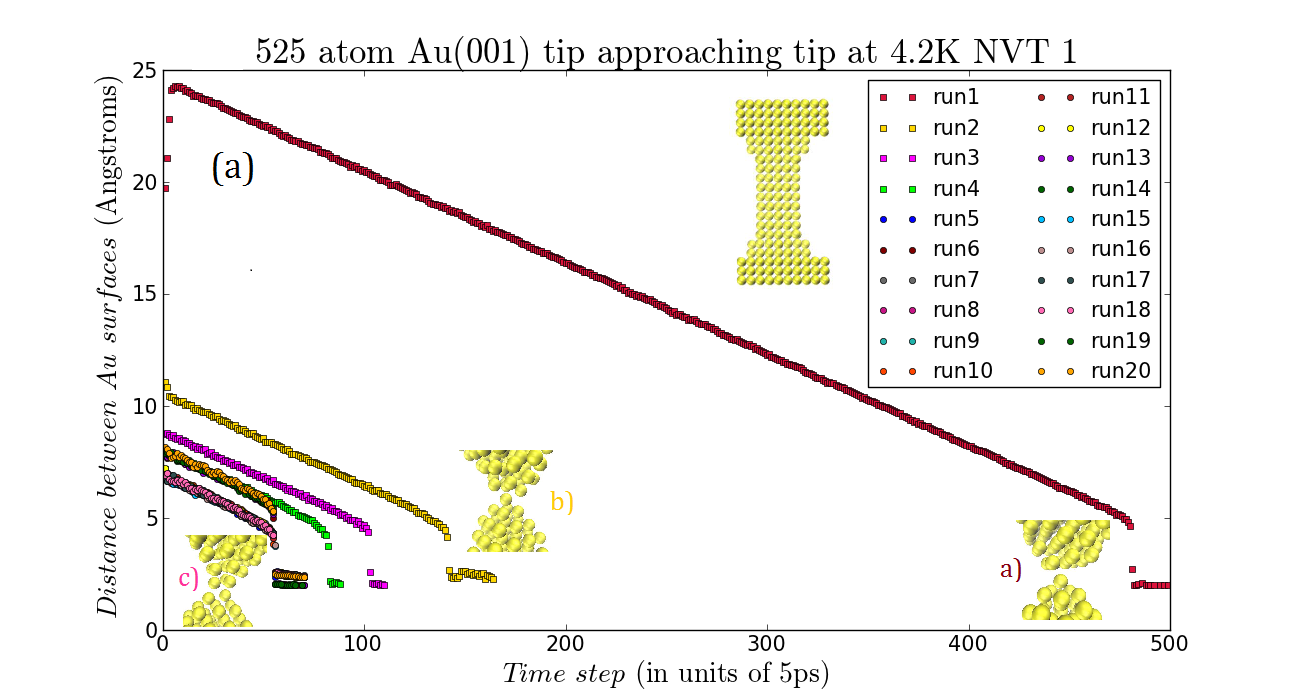}
\includegraphics[width=0.49\textwidth]{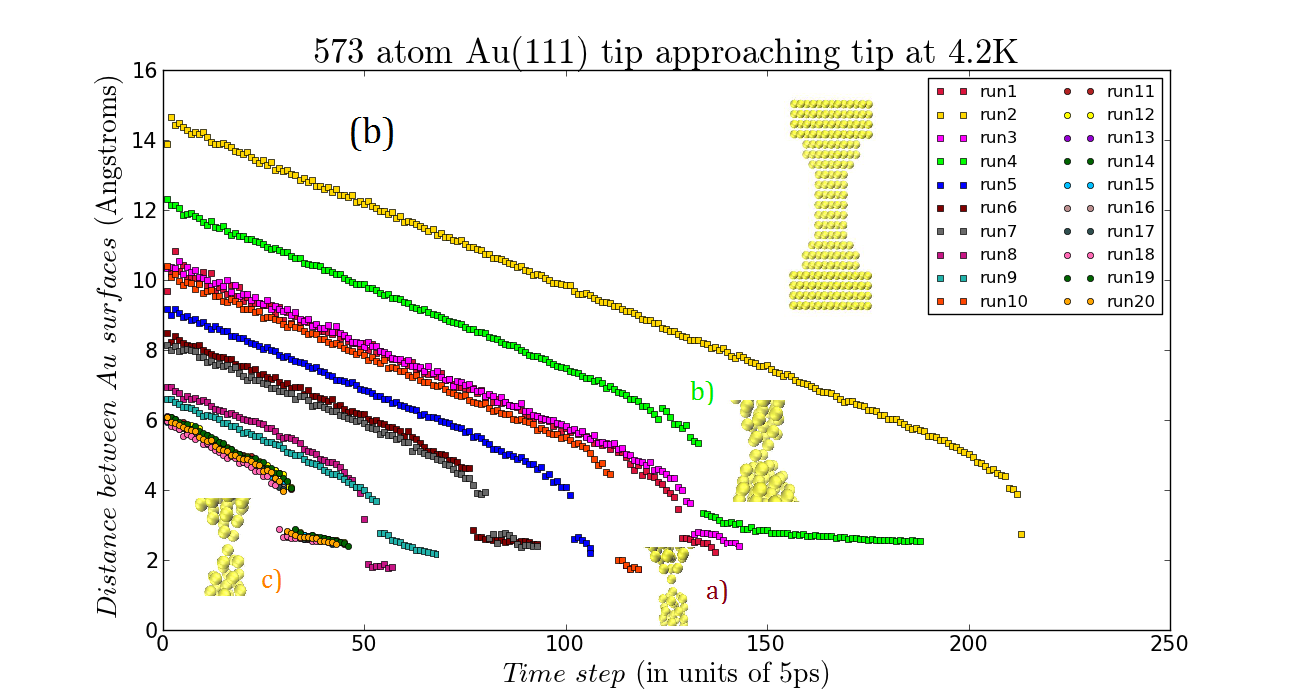}

\includegraphics[width=0.49\textwidth]{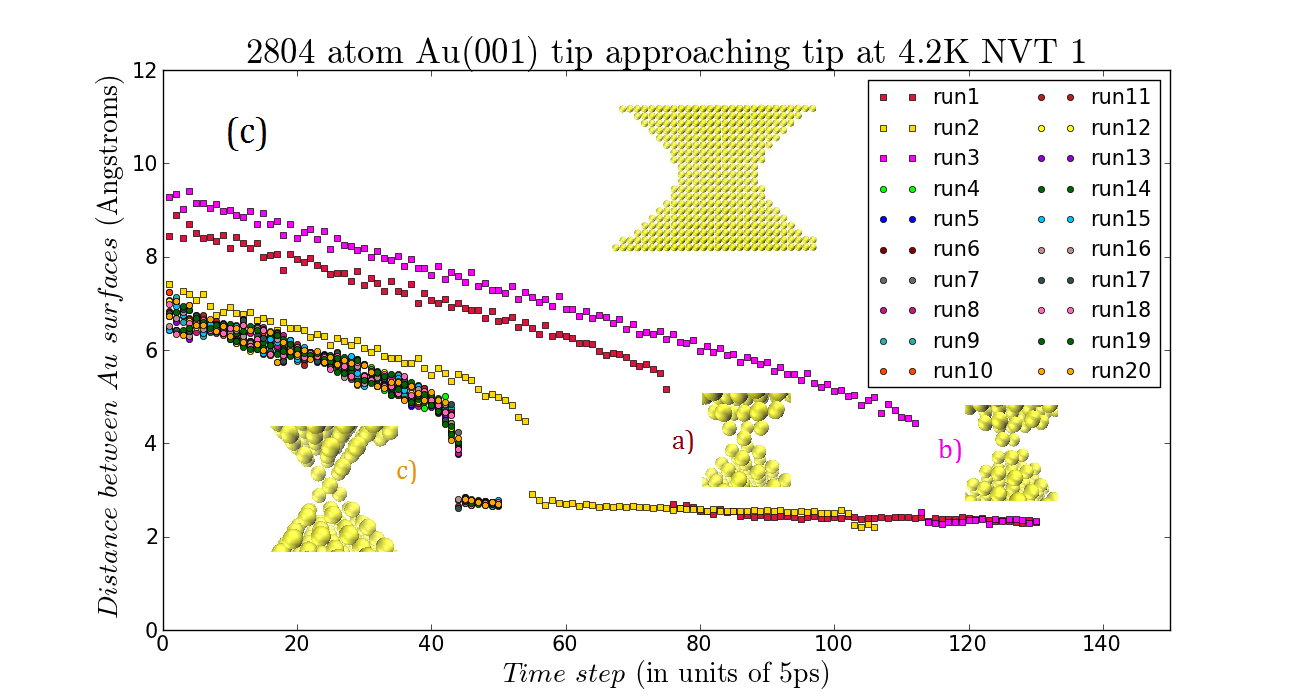}
\includegraphics[width=0.49\textwidth]{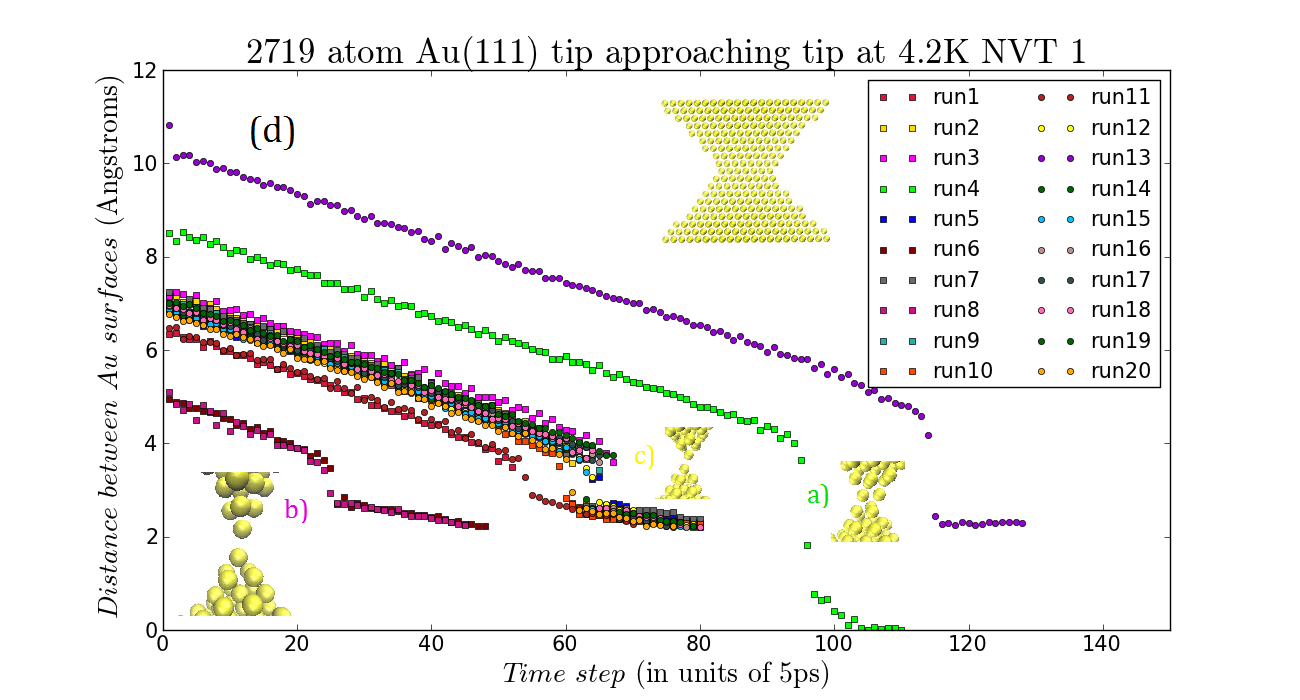}
\caption{(Color online) Traces of distance between electrodes versus time for a system of: (a) 525 atoms oriented along the (001) direction. Insets a, b and c correspond to the geometries of the electrodes at the moment of first contact for formation cycles labelled run1, run2 and run18, respectively; (b) 573 atoms along (111). Insets a, b and c correspond to the first-contact geometries of formation cycles run1, run4 and run20, respectively; (c) 2804 atoms along (001). Insets a, b and c correspond to the first-contact geometries of formation cycles run1, run3 and run20, respectively; (d) 2719 atoms along (111). Insets a, b and c correspond to the first-contact geometries of formation cycles run4, run8 and run12, respectively.}
\label{fig3}
\end{figure}

Figures \ref{fig3}(a), (b), (c) and (d) show the MD formation traces of the four structures considered here. In all cases we observe that the first few cycles of formation (labelled "run1", "run2", etc.) require more steps before first contact, with the exception of "run13" in Fig. \ref{fig3}(c). All the traces clearly exhibit a jump in distance at the moment of first contact between the electrodes. This jump is reminiscent of those observed in experimental conductance traces \cite{UntiedtJC}. We also see that after a number of cycles the traces all overlap, indicating that the geometric configuration of the contact repeats in a regular fashion.  When the traces repeat in this way we can talk about a process of mechanical annealing or, in other words, that we now have reproducible contacts with sharpened tips.

Traces 1 to 6 in Fig. \ref{fig3}(a) (525 atoms, (001) orientation) differ significantly from each other: the jump occurs at different time steps, and its height also differs from trace to trace. However, after run 7 all traces up to run 20 are nearly the same: the jumps occur at the same time step and are nearly equal in height. The insets in Fig. \ref{fig3}(a) show the atomic configurations right before contact for several traces. In many cases contact is made through a single atom, or \emph{monomer}; in other cases, as in inset c, contact is made through two adjacent atoms, or \emph{double contact}. There is greater diversity among the traces in Fig. \ref{fig3}(b) (573 atoms, (111) orientation), both in terms of the time taken for the jump to occur and its height. Here, it takes at least 10 indentation cycles before the traces start to overlap. When the traces are stable, the jump to contact occurs through a monomer (inset c). However, there are also cases where the jump occurs through a \emph{dimer} (inset a) and more than one atom (inset b). In Fig. \ref{fig3}(c) (2804 atoms, (001)), the traces overlap after only three cycles. Note that the aspect ratio of this neck (the length of the neck divided by the diameter of its minimum cross-section) is lower ($\sim 2:1$) than that of the structure in Fig. \ref{fig3}(a) ($\sim 5:1$). As in Fig. \ref{fig3}(b), the traces in Fig. \ref{fig3}(d) (2719 atoms, (111)) exhibit greater diversity. The most energetically stable surface layers of a face-centred cubic (FCC) crystal are oriented along the (111) crystallographic direction, since atoms in the exposed layers have more nearest neighbours than in any other direction. Because (111) layers form on the oblique faces of the pyramid shaped electrodes that are oriented along (001), e.g. in Figs. \ref{fig3}(a) and \ref{fig3}(c), it is easier to "cold" anneal surfaces in this direction, requiring less than half the number of cycles than is needed to achieve reproducible structures along (111). The insets in Fig. \ref{fig3}(d) show that contact occurs more often through a vertical dimer than through a monomer, and double atomic contacts also occur.

\subsection{Results of the DFT calculations on the MD-generated contacts}
Following Refs. \cite{UntiedtJC,Undersab}, we have selected three representative (111) contact geometries from the MD atomic configurations at the moment of first contact: the monomer in inset b and the double contact in inset c in Fig. 3(b) and the dimer in inset c in Fig. 3(d). We have calculated the conductance of these structures since the conductance of their counterparts along (001) have already been calculated elsewhere \cite{Undersab}. The results are shown in Fig. 4 and the conductance values are expressed in units of $G_0$. The crystallographic orientation and cycle of formation of the three configurations are also indicated, as well as the time step of and distance to first contact.

\begin{figure}[htp!]
\centering 
\begin{minipage}{18pc}
\includegraphics[width=\textwidth]{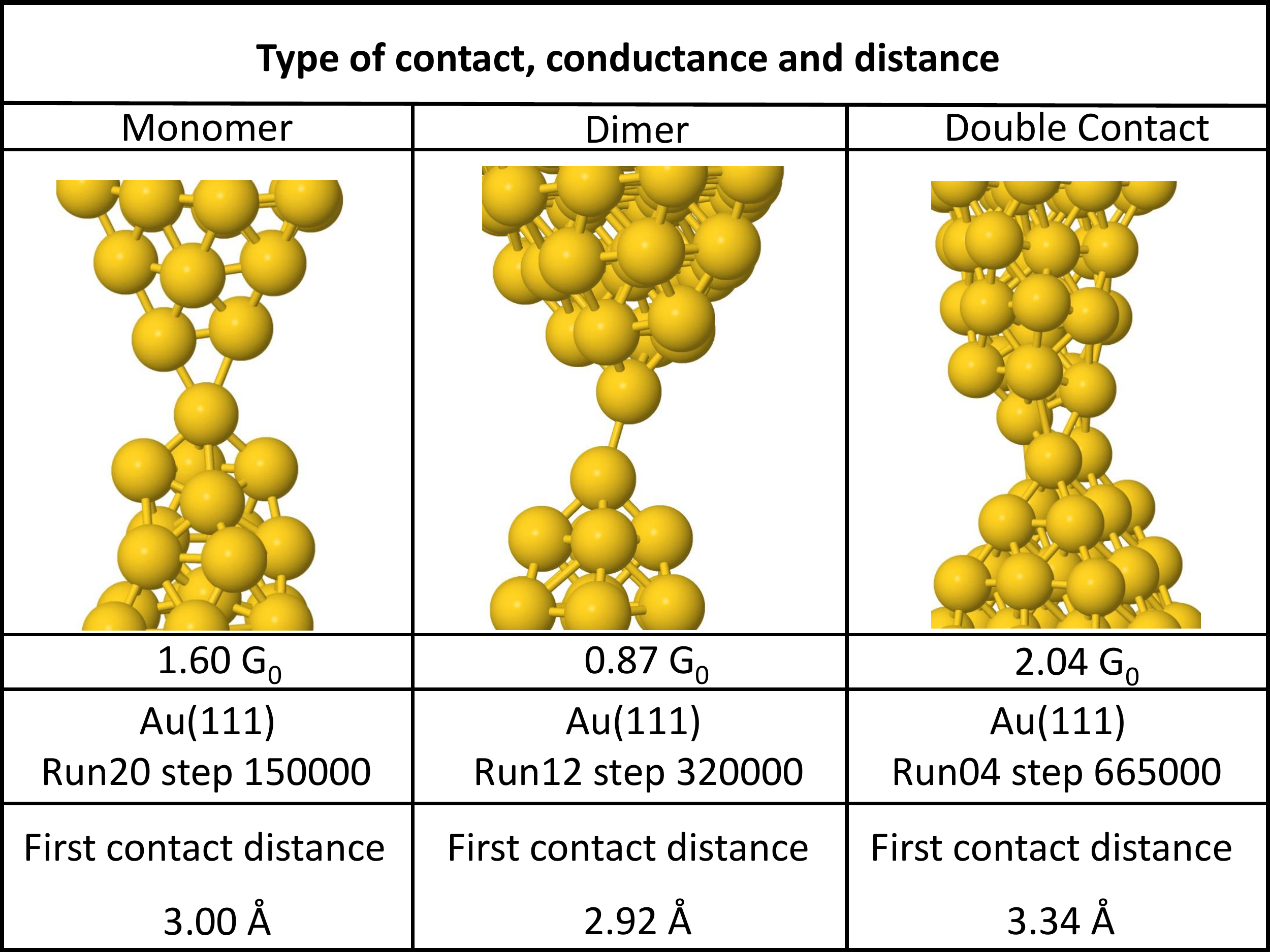}
\end{minipage}\hspace{1.5pc}
\begin{minipage}[b]{18pc}
\caption{(Colour online) Different types of contact: monomer, dimer and double contact and their corresponding conductances. The distance to first contact in each case is also shown. \label{fig4}}
\end{minipage}
\end{figure}

 We should expect the conductance of the monomer to be between that of the dimer and the double contact, since the tunnelling current across atoms just above and below the monomer ought to be higher than the dimer's -- because of the closer proximity of the atoms around the monomer. It appears that neighbouring atoms in the minimum cross-section make a substantial contribution to conductance in the form of tunnelling current, which could explain the high value of 1.60 $G_{0}$ for the monomer. The size of the distance to first contact is also consistent with the type of contact under consideration. Thus, one would expect the jump to be smallest in the case of the dimer -- the approach of two single atoms towards each other -- since the oncoming atoms "see" fewer atoms, and neighbours on the same electrode effectively keep them from jumping to the opposing electrode for longer than would be expected. This contrasts with the case of a single atom feeling the attractive force of more than one atom on the opposing tip, e.g., the monomer in Fig. 4. The force of attraction between more than one atom on both opposing electrodes, e.g., the double contact in Fig. 4, ought to be the greatest and hence so should the jump to contact, as is indeed the case.

\section{Conclusion}

 We have shown previously \cite{MecSab,youtubemecansab} that when repeated indentation is performed on structures that are oriented in the (001) direction, traces of \emph{minimum cross-section} versus time overlap perfectly. This has been explained as a process of mechanical annealing of the electrodes that results in very sharp and stable tips \cite{MecSab}. Here we observe the same behaviour, repetition of
the traces after several cycles of indentation, but of the \emph{minimum distance} between the electrodes vs time. Moreover, we show that electrodes oriented along (111) can also be mechanically annealed.  In contrast to (001) electrodes, the traces of (111) electrodes take longer to repeat in a regular way, apparently due to the stability of the exposed (111) surface layers on the oblique faces of (001) structures. In the case of (111) electrodes we also find that the final jump to contact occurs predominantly through a monomer or a dimer, and to a lesser extent, through a double contact.

The conductance values we obtain are in agreement with previous works \cite{Undersab}. However, in the case of the monomer it is higher than expected. Instead of using a minimal one-electron basis set, it might be necessary to use an eleven-electron basis set to improve this result \cite{MecSab}.

\ack

In compiling this paper, W. Dednam has made extensive use of python scripts written by Prof Andr\'{e} E. Botha, who kindly shared them. W. Dednam acknowledges support from the National Research Foundation of South Africa through the Scarce Skills Masters scholarship funding programme. This work is supported by the Generalitat Valenciana through grant reference PROMETEO2012/011 and the Spanish government through grant FIS2010-21883. This work is part of the research programme of the Foundation for Fundamental Research on Matter (FOM), which is financially supported by the Netherlands Organisation for Scientific Research (NWO).
 

\section*{References}
\bibliographystyle{iopart-num}
\bibliography{nanocontacts}
 
\end{document}